\documentclass[11pt]{article}
\usepackage{amsfonts,amssymb}
\usepackage{epsfig}
\usepackage{color}
\usepackage{graphicx}

\newtheorem{theorem}{Theorem}{\bf}{\it}

\usepackage{epsfig}
\usepackage{latexsym}
\usepackage{amsmath}
\usepackage{mathbbol}
\renewcommand{\forall}{\mbox{for all}\,\,}

\def\eps{\varepsilon}
\def\C{{\cal C}}

\def\Halmos{\quad\hfill$\Box$}
\def\integers{{\mathbb Z}}        
\def\naturals{{\mathbb N}}

\def\reals{{\mathbb R}}

          \title{Spin \& Statistics\\ in Nonrelativistic Quantum
            Mechanics, II}
          \author{Bernd Kuckert\\II. Institut f\"ur Theoretische Physik, Universit\"at Hamburg\\
            Luruper Chaussee 149, 22761 Hamburg, Germany\\
\vspace{0.5mm}\\Jens Mund\\
Instituto de F\'isica, Universidade de S\~ao Paulo\\
CP 66.318, 05315-970, S\~ao Paulo, SP Brasil}

\begin{document}
\maketitle

\begin{abstract}
  Recently a sufficient and necessary condition for Pauli's
  spin-sta\-tis\-tics connection in nonrelativistic quantum mechanics
  has been established \cite{Kuc04}. The two-dimensional part of this
  result is extended to $n$-particle systems and reformulated
  and further simplified in a more geometric language.
\end{abstract}

\noindent%
A nonrelativistic quantum system in two spatial dimensions satisfies
Pauli's spin-statistics law if and only if the total (i.e., spin plus
orbital) angular momentum of the two-particle system with respect to
its center of mass is twice that of each of the particles
\cite{Kuc04}. In what follows, we reformulate this result in a more
geometric language, which should further clarify and simplify notions,
results, and proofs. For a discussion of the most relevant literature
and the corresponding references, see Ref. \ref{Kuc04}.

Following Laidlaw and DeWitt \cite{LD}, the configuration space of $n$
distinguishable particles in $\reals^2$ is described by the set $Y_n$
of all $n$-tuples of points in the plane no two of which coincide:
$$Y_n:=\{y=({\bf x}_1,\dots,{\bf x}_n)\in(\reals^2)^n:\,
{\bf x}_i\neq {\bf x}_j\,{\mbox{for}}\,i\neq j\}.$$
A  right  action of the symmetric group $S_n$ on this space is defined by
$$y\pi:=({\bf x}_{\pi(1)},\dots,{\bf
  x}_{\pi(n)}),\quad\pi\in S_n.$$
The orbits of $S_n$ in $Y_n$
yield the configuration space $X_n:=Y_n/S_n$ of $n$ indistinguishable
particles. 

Recall that the fundamental group of $X_n$ is the braid group $B_n$
and that $B_n$ acts on the universal covering space $\tilde
X_n\stackrel{\Lambda}{\to}X_n$ from the right. Namely, choose any base
point $x_0\in X_n$; then each $x\in \tilde X_n$ can be represented as
the homotopy equivalence class $[\xi]$ of any curve $\xi$ joining
$x_0$ to a point $\Lambda(x)\in X_n$. The map $\Lambda$ defined this
way is the covering map from $\tilde X_n$ onto $X_n$.  The set
$B_n=\Lambda^{-1}(\{x_0\})$ of all homotopy classes of closed curves
starting and ending at $x_0$ is endowed with a group structure by
concatenation.  Concatenation of closed curves at $x_0$ with curves
from $x_0$ to $\Lambda(x)$ then defines a right action $(x,b)\mapsto
xb$ of $B_n$ on $\tilde X_n$.

In two-dimensional systems, the wave functions of indistinguishable
particles are, in general, not ``single-valued'' under an exchange of
the particles, i.e., the phase they pick up along a curve in $X_n$
exchanging the particles depends on the homotopy class of the curve.
Mathematically, this means that the pure-state space of $n$
indistinguishable particles is not described by the square-integrable
functions on $X_n$.  Instead, the statistics of the particle species
under consideration is characterized by a 
unitary representation $\eps$ of $B_n$. In what follows, $\eps$ will
be assumed to be scalar, i.e., to take values in $U(1)$. The 
wave functions, then, are regular
distributions on the universal covering space $\tilde{X_n}$ that are
{\it equivariant} with respect to $\eps$, i.e.,
$\Psi(xb^{-1})=\eps(b)\Psi(x)$.  As this implies
$|\Psi(x)|^2=|\Psi(xb^{-1})|^2$, a probability density function on the
configuration space $X_n$ is obtained eventually, and this density is,
as usual, required to induce a finite measure.

We denote the space of wave functions by $L^2_\eps(\tilde{X_n})$. If 
$\tau_1,\dots,\tau_{n-1}$ denote the transpositions generating $B_n$,
then one concludes from the braid relation
$\tau_i\tau_{i+1}\tau_i=\tau_{i+1}\tau_i\tau_{i+1}$, $1\leq i\leq
n-2$, that
$\eps(\tau_1)=\dots=\eps(\tau_{n-1})=:\kappa$. This phase is
the statistics phase of the particle species; it equals $1$ for bosons
and $-1$ for fermions, whereas it can equal any phase for anyons. In
what follows, we write $L^2_\kappa$ instead of $L^2_\eps$,  as  
$\kappa$ fixes $\eps$.

For each $\alpha\in\reals$ and each $n\in\naturals$, let $d_n(\alpha)$
denote the counterclockwise rotation by the angle $\alpha$ acting on
$X_n$, and let $\tilde d_n$ denote the unique lift of the action $d_n$
to an action on $\tilde X_n$. The universal covering group
$(\reals,+)$ of $SO(2)$ acts on the $n$-particle wave functions
 $L^2_\kappa(\tilde X_n)$  by
 $$D_n(\theta)\Psi(x)=e^{i{{s}}_n\theta}\Psi(\tilde d_n(-\theta)x),\quad
 x\in \tilde{X_n};$$
where $s_n\in\reals$.

$X_2$ factorizes into a center-of-mass coordinate in $\reals^2$ and a
relative coordinate in the cone $\C:=(\reals^2\backslash\{0\})/\{x=
-x\}$, and $\tilde{X_2}$ factorizes into a center-of-mass coordinate
in $\reals^2$ and a relative coordinate in $\tilde\C$.  As a
consequence, the two-particle Hilbert space is the tensor product of
the center-of-mass Hilbert space $L^2(\reals^2)$ and the
relative-coordinate Hilbert space $L^2_\kappa(\tilde{\C})$.  Further,
the unitary action $D_2$ factorizes in such
a way that on $L^2_\kappa(\tilde\C)$ it acts as
$$D_2^{\rm rel}(\theta)\Psi(r,\chi):=\Psi(r,\chi-\theta),\quad
\theta\in\reals,\,(r,\chi)\in\tilde\C.$$
Here we have identified
$\tilde\C$ with $\reals^{>0}\times\reals$. Since the braid $\tau_1$ acts as
$(r,\phi)\tau_1=(r,\phi+\pi)$, it follows that $D_2^{\rm
  rel}(\pi)=\kappa$.

\bigskip

\begin{theorem}
  Pauli's spin-statistics connection $\kappa=e^{2\pi i{s}}$ holds if
  and only if there exists a unitary operator $V:L^2(\reals^2)\to
  L^2_\kappa(\tilde\C)$ such that
$$D_2^{\rm rel}(\theta)V=VD_1(2\theta)\qquad\forall\, \theta\in\reals.$$
\end{theorem}

{\it Proof.}  The condition implies $\kappa=D^{\rm
  rel}_2(\pi)=VD_1(2\pi)V^*=e^{2\pi i{s}}$, so it is sufficient.
Conversely, assume $\kappa=e^{2\pi i{s}}$, and define
$\Lambda(\xi):=\xi-2\pi[\xi/(2\pi)]$, where $[\,\cdot\,]$ is the
Gau\ss\,bracket. Using polar coordinates in $\reals^2$ and the
assumed spin-statistics connection, define
$V: L^2(\reals^2)\to L^2_\kappa(\tilde\C)$ by
$$V\Psi:=[\underbrace{(r,\chi)}_{\in\reals^{>0}\times\reals}]
\mapsto
  e^{-2i{s}\chi}\Psi(r,\Lambda(2\chi))].$$
One obtains
\begin{align*}
  D_1(-2\theta)V^*D_2^{\rm rel}(\theta)V\Psi&=D_1(-2\theta)V^*
  D_2^{\rm rel}(\theta)[(r,\chi)\mapsto
  e^{-2i{s}\chi}\Psi(r,\Lambda(2\chi))]\\
  &=D_1(-2\theta)V^*[(r,\chi)\mapsto 
    e^{-2i{s}(\chi-\theta)}\Psi(r,\Lambda(2(\chi-\theta)))]\\
  &=D_1(-2\theta)[\underbrace{(r,\phi)}_{\in\reals^{>0}\times[0,2\pi)}\mapsto
  e^{2i{s}\theta}\Psi(r,\Lambda(\phi-2\theta))]=\Psi.
\end{align*}
\Halmos

The generator of $D_2^{\rm rel}$ represents the total angular momentum
of the two-particle system with respect to its center of mass. The
generator of $D_1(2\,\cdot)$ represents {\it twice} the total angular momentum
operator of the single particle.
Pauli's spin-statistics connection, then, is equivalent to the
condition that these operators are unitarily equivalent.

The statistics phase $\kappa_n$ of an $n$-particle system equals
$\kappa^{n^2}$, and if an $n$-particle system is rotated by
$2\pi$,
it picks up a phase $\kappa^{n(n-1)}$ due to the equivariance
property, i.e.\ $\Psi(\tilde d_n(-2\pi)x)=\kappa^{n(n-1)}\Psi(x)$. All
this can be shown by straightforward braid diagrammatics; see the
following diagrams for the case $n=3$.

\bigskip
\begin{center}

\epsfig{file=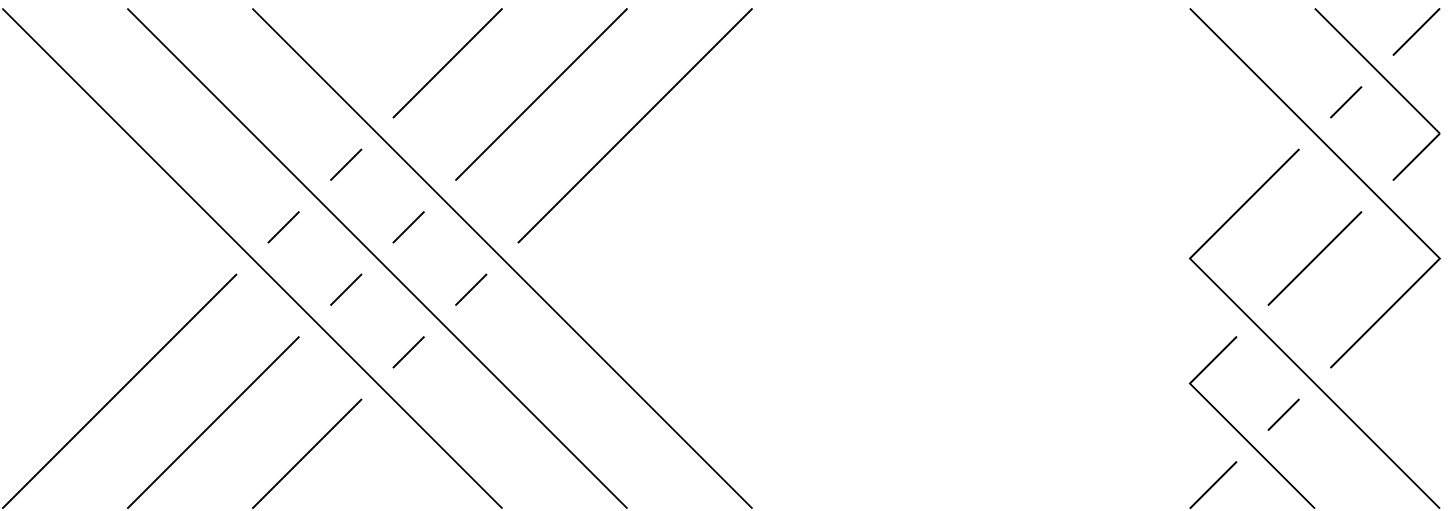,width=12cm}

\end{center}
(A) Exchange: $9=3^2$ crossings.\hfill(B) Rotation:
$6=3(3-1)$ crossings
\vspace{0.3cm}

\begin{theorem}
  If a single particle satisfies the spin-statistics relation
  $e^{2\pi i s}=\kappa$, then a system of $n$
  indistinguishable copies of this particle satisfies the
  spin-statistics relation $D_n(2\pi)=\kappa_n$ 
  if and only if ${{s}}_n\in ns+\integers$.
\end{theorem}

{\it Proof.} Since $D_n(2\pi)=e^{2\pi i{{s}}_n}\kappa^{n(n-1)}$, one has
$D_n(2\pi)=\kappa_n$ if and only if $e^{2\pi
  i{{s}}_n}\kappa^{n(n-1)}=\kappa^{n^2}$, i.e., $e^{2\pi
  i{{s}}_n}=\kappa^n=e^{2\pi in s}$. \Halmos


{\it Acknowledgements.} B. K. has been supported by the
Emmy-Noether programme of the Deutsche Forschungsgemeinschaft.
J.M. has been supported by the Brazilian foundation
FAPESP.

\end{document}